\documentclass[twocolumn]{openjournal}
\pdfoutput=1 

\usepackage{natbib}
\usepackage{graphicx}
\usepackage{amsmath}

\usepackage{rotating}

\newcommand{\spm}[2]{\ensuremath{^{+#1}_{-#2}}}

\begin{document}

\journalinfo{The Open Journal of Astrophysics}
\shorttitle{A Pilot ZTF Self-Lensing Search}
\shortauthors{Crossland et al.}

\title{A Pilot Search for Gravitational Self-Lensing Binaries\\ with the Zwicky Transient Facility}

\author{Allison Crossland$^{1,*}$}
\author{Eric C. Bellm$^{1,\dagger}$}
\author{Courtney Klein$^2$}
\author{James R. A. Davenport$^1$}
\author{Thomas Kupfer$^3$}
\author{Steven L. Groom$^4$}
\author{Russ R. Laher$^4$}
\author{Reed Riddle$^5$}

\affiliation{$^1$ DIRAC Institute, Department of Astronomy, University of Washington, 3910 15th Avenue NE, Seattle, WA 98195, USA}
\affiliation{$^2$ Department of Physics and Astronomy, University of California Irvine, CA 92697, USA}
\affiliation{$^3$ Hamburg Observatory, University of Hamburg, Gojenbergsweg 112, 21029 Hamburg, Germany}
\affiliation{$^4$ IPAC, California Institute of Technology, 1200 E. California Blvd, Pasadena, CA 91125, USA}
\affiliation{$^5$ Caltech Optical Observatories, California Institute of Technology, 1200 E. California Blvd, Pasadena, CA 91125, USA}

\email{$^*$ allifc@uw.edu}
\email{$^\dagger$ ecbellm@uw.edu}

\begin{abstract}
Binary systems containing a compact object may exhibit periodic brightening episodes due to gravitational lensing as the compact object transits the companion star. Such ``self-lensing'' signatures have been detected before for white dwarf binaries. 
We attempt to use these signatures to identify detached stellar-mass neutron star and black hole binaries using data from the Zwicky Transient Facility (ZTF).
We present a systematic search for self-lensing signals in Galactic binaries from a subset of high-cadence ZTF data taken in 2018. We identify 12 plausible candidates from the search, although because each candidate is observed to only brighten once, other origins such as stellar flares are more likely. 
We discuss prospects for more comprehensive future searches of the ZTF data.
\end{abstract}

\maketitle




\section{Introduction}\label{sec:Introduction}

The observed population of Galactic compact objects is shaped by the methods by which they are discovered.
For binary systems with active mass transfer, instabilities in the accretion disk or wind accretion can create high X-ray luminosities detectable by all-sky monitors \citep[for a recent review, see][]{Kalemci:22:BHSpectralTimingReview}. 
Detached, non-interacting compact objects are expected to be more numerous, but harder to detect. 
Large radial velocity shifts may be identified in spectroscopic surveys \citep[e.g.,][]{Thompson:19:NoninteractingBH,Yi:22:RVNSbinary,Shenar:22:XrayQuietBHLMC}, although other stellar binaries may mimic these signatures \citep[][and references therein]{Bodensteiner:22:BHImpostersReview}.
\textit{Gaia} has identified compact objects through their astrometric wobble \citep{El-Badry:23:GaiaBH1,El-Badry:23:GaiaBH2}.
And microlensing can identify free-floating compact objects as the gravity of the compact object bends the light of the background star towards the observer \citep{Sahu:22:MicrolensedBH, Lam:22:MicrolensedMassGap}.

For edge-on binary systems containing a compact object and a companion star, we may see a periodic brightening pulse due to gravitational ``self-lensing'' as the compact object transits the companion \citep{Maeder:73:SelfLensing,Gould:95:SelfLensing}.
Multiple stellar-mass self-lensing systems have been discovered with \textit{Kepler} \citep{Kruse:14:SelfLensingWD, 2018AJ....155..144K, Masuda:19:SelfLensingWD}\footnote{Some authors have interpreted periodic temporal signatures from the supermassive black hole systems in AGN as self-lensing \citep{DOrazio:18:AGNSL,Hu:20:Spikey,Ingram:21:QPESelfLensing}.},  all of which to date contain a white dwarf (WD).
In some eclipsing WD systems, lensing has been suggested to modify the observed eclipse depth \citep[e.g.,][]{Steinfadt:10:Discovery-of-the-Eclipsing-Detached,Muirhead:13:Characterizing-the-Cool-KOIs,Kaplan:14:Properties-of-an-Eclipsing-Double}, although the statistical significance has been generally low.

However, population synthesis estimates suggest that thousands of neutron star (NS) and black hole (BH) binaries could have detectable self-lensing signals \citep{Wiktorowicz:21:SelfLensing}. 
As the self-lensing signal from NS and BH systems is higher amplitude than from WDs, it may be detectable at the photometric precison available from ground-based synoptic surveys.
Such surveys provide an advantage of greater depth and temporal duration, allowing a search for self-lensing signals from many more stars than can be observed by \textit{Kepler} or \textit{TESS}.

The Zwicky Transient Facility \citep[ZTF;][]{Bellm:19:ZTFOverview,Graham:19:ZTFScience} provides a powerful dataset for such a search. 
ZTF is a wide-field time-domain survey of the Northern Hemisphere sky that uses a mosaic CCD camera \citep{Dekany:20:ZTFObservingSystem} mounted on the 48-inch Samuel Oschin Schmidt Telescope at Palomar Observatory. 
ZTF’s high observing cadence (minutes--days), large field of view (47\,deg$^2$), and moderate single-image depth ($m_{\rm lim} \sim 20.5$\,mag) provide an opportunity to detect relatively rare and short-lived self-lensing pulses \citep{Wiktorowicz:21:SelfLensing, Nir:23:WDSelfLensing}.


In this paper, we describe a pilot search for NS and BH self-lensing in a small subset of continuous cadence ZTF data taken in the Galactic Plane.
We describe the ZTF dataset used for this search, outline our process of searching for self-lensing pulses, and present final candidates from the search (Table \ref{table:masses}). We close by discussing the interpretation of these candidates as well as prospects for future self-lensing searches with ZTF.

\section{Methods and Data} \label{sec:Methods}

\subsection{Dataset} \label{sec:Data}


The Zwicky Transient Facility is a high-cadence, wide area optical time domain survey.
ZTF has conducted a variety of surveys to date \citep{Bellm:19:ZTFScheduler}.
Since we expect the duration of self-lensing pulses to be on the order of minutes to hours, for our initial search we focused on a subset of ``continuous cadence'' data collected during August of 2018 \citep{Kupfer:21:ZTFHCYear1}.
The goal of these observations was to identify short-timescale variability through near-continuous cadence observations of a small set of high stellar density fields.
In the August 2018 campaign, fourteen ZTF fields were observed.
Fields were observed in pairs, with each pair observed for two nights.
During observations, ZTF generally alternated between the two fields over a period of about three hours, taking 30 second exposures at each position in the ZTF $r$ band.

The images were reduced by ZTF's automated data processing pipelines \citep{Masci:19:ZTFDataSystem}.
The ZTF data system produces PSF photometry catalogs for each direct (unsubtracted) image; these catalogs are then matched to deeper source catalogs from the coadded ZTF reference image to produce object light curves, grouped by the ZTF field, readout channel, and filter.
From the 14 fields we have light curves for 117,483,764 sources.
We performed analysis both on HDF5 ``matchfiles'' produced by the IPAC pipelines as well as in a version ingested in the Astronomical eXtensions for Spark (AXS) system \citep{Zecevic:19:AXS}.

\subsection{Candidate Detection}
\label{sec:cand}

Our initial categorization of these data was motivated by a search for stellar flares \citep{Klein:20:ZTFFlaresAAS}. 
As the search algorithm can identify a range of outburst shapes, we recognized that the raw flare candidates might also include self-lensing signals.

The initial flare candidates were identified using the {\tt FINDflare} change-point algorithm \citep{Chang:15:FindFlares}.
We defined a continuous light curve as a series of observations with a gap between observations no larger than 30 minutes. 
We treated each continuous section of the light curve separately and disregarded any epoch with a ZTF \texttt{catflag} $>$ 0. 
Our average light curve duration was 2.35 hours. Due to short time scales observed, we did not worry about the effects of periodic variability.

To detect spikes in flux we used the \texttt{FINDflare} criteria of:
\begin{equation}
x_i - x_L > 0
\end{equation}
\begin{equation}
\frac{|x_i - x_L|}{\sigma_L} > 3
\end{equation}
\begin{equation}
\frac{|x_i - x_L + w_i|}{\sigma_L} > 1
\end{equation}
where $x_i$ is the {\tt psfflux}  of a given epoch, $x_L$ is the median value of flux in the continuous light curve, $\sigma_L$ is the standard deviation of the flux in the continuous light curve, and $w_i$ is the {\tt psffluxerr} of a given epoch. When three consecutive observations meet these criteria it was counted as a flare candidate. This search produced 15,262 flare candidates. Due to the duration of the continuous-cadence light curve segments we were limited to identifying one flare per segment.
We did not impose any selection criteria based on position in the color-magnitude diagram or detection by \textit{Gaia}.


\subsection{Temporal Profile Modeling}
\label{sec:model}

To distinguish stellar flares from candidate self-lensing pulses, we used nonlinear least-squares to fit the temporal profile of the candidates identified by \texttt{findFlare} with three analytic models.
We fit each lightcurve with a fast-rise, exponential decay stellar flare model \citep{Daven_2014}, a Gaussian, and a constant.
We used the Akaike information criterion (AIC) to select lightcurves which were best-fit by the Gaussian model \citep{1974ITAC...19..716A}. Given the AIC values of the flare and Gaussian models for each source, the likelihood that the Gaussian model minimizes information loss relative to the flare model can be found by comparing the AIC values:

\begin{equation}
\exp\left({\frac{\mathrm{AIC}_\mathrm{gauss}-\mathrm{AIC}_\mathrm{flare}}{2}}\right) < e^{-2}.
\end{equation}

\begin{figure}
    \centering
    \includegraphics[width=0.45\textwidth]{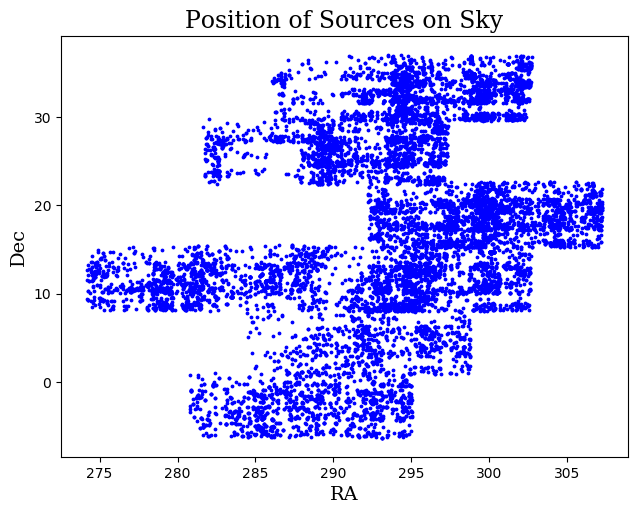}
    \caption{RA/Dec positions of 1910 identified bright pulses in our 2018 continuous-cadence ZTF dataset.}
    \label{fig:positions}
\end{figure}

Our chosen cutoff ($e^{-2}$) is comparable to a 95\% confidence selection.
This approach yielded 1910 instances brightening pulses better fit by a symmetric, Gaussian-like profile that we investigated as potential self-lensing signals (Figure \ref{fig:positions}). 

\subsection{Initial Screening} \label{sec:Initial Screening}

We visually inspected each of the 1910 continuous-cadence light curve segments with identified bright pulses. We identified three broad categories: noisy data that clearly did not exhibit the desired self-lensing pulse shape, remaining curves with asymmetric fast-rise, exponential decay shapes characteristic of stellar flares but not excluded by our model selection procedure (\S \ref{sec:model}), and outbursts of interest that we analyzed further. Approximately 90\% of the 1910 candidates fell into the first two categories and were discarded. 

\begin{figure}
    \centering
    \includegraphics[width=0.45\textwidth]{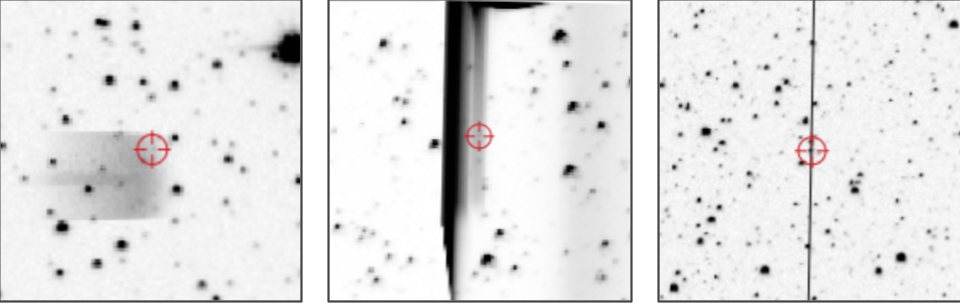}
    \caption{Three examples of artifacts found in our dataset that caused spurious brightening pulses.}
    \label{fig:artifacts}
\end{figure}

Next, using the NASA/IPAC Infrared Science Archive (IRSA), we more carefully inspected the images of each source at the time of the potential pulse. Around 80 instances of data artifacts were discovered (e.g., Figure \ref{fig:artifacts}), and those candidates were rejected.

\begin{figure}
    \centering
    \includegraphics[width=0.45\textwidth]{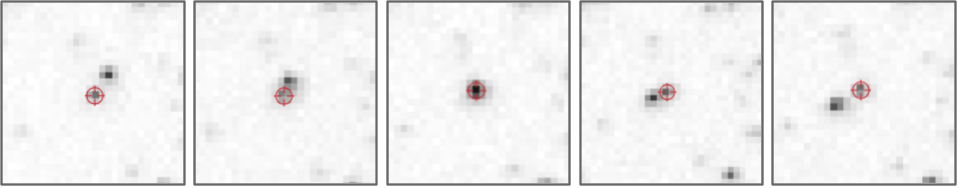}
    \caption{Consecutive images showing an asteroid transiting a star (circled in orange).}
    \label{fig:asteroid}
\end{figure}

Asteroids transiting a background star were another frequent contaminant (Figure \ref{fig:asteroid}; see also \citealp{Sorabella:23:TESSSLAsteroids}). 
As an asteroid passes in front of a background star, we see a short, symmetrical brightening pulse, similar to what we might expect of a self-lensing pulse. Using SkyBot \citep{2006ASPC..351..367B}, a solar system body identification service, and the python package Astroquery \citep{2019AJ....157...98G}, we cross-matched the time and location of each pulse in our dataset to all known asteroid positions in the solar system. We looked for instances of an asteroid passing within 15 arcseconds of the source location at the same time as the brightening pulse. We rejected 50 candidates coincident with an asteroid transit.

Finally, to rule out any brightening pulses caused by hidden data artifacts, we compared each of the remaining light curves to that of a neighboring, non-variable star.
A neighboring source exhibiting a similar pulse as our target indicated that the pulse was caused by an artifact in the data processing or calibration, not self-lensing. Four instances of neighbor variability were found, and those candidates were rejected. 

After these steps, we were left with 23 self-lensing candidates.

\subsection{Mass Estimation}\label{sec:mass_estimation}

Given a self-lensing signal, it is possible to estimate the mass of the compact object.
We used the approximate relations of \citet{Masuda:19:SelfLensingWD} for the self-lensing pulse amplitude 

\begin{equation}
\begin{split}
    s_{sl}=2 \left (\frac{R_E}{R_\star} \right )^2
    = 
    7.15 \times 10^{-5} \left (\frac{R_\star}{R_\odot} \right )^{-2}
    \left (\frac{P}{1 day} \right )^{2/3} \\
    \times \left (\frac{M_\bullet}{M_\odot} \right )
    \left (\frac{M_\bullet +M_\star}{M_\odot}\right)^{1/3}
\end{split}
\end{equation}
    

and duration

\begin{equation}
\begin{split}
    \tau_{sl} = \frac{R_\star P}{\pi a} \cdot \frac{\pi}{4} = 
    1.8\, \mathrm{hr} \times \frac{\pi}{4}\left(\frac{P}{1 day}\right)^{1/3} \\
    \times \left(\frac{M_\bullet +M_\star}{M_\odot}\right)^{-1/3} \left(\frac{R_\star}{R_\odot}\right)
\end{split}
\end{equation}

where $M_\bullet$ is the mass of the compact object and $M_\star, R_\star$ the mass and radius of the companion star.
These approximations are valid when $R_E \ll R_\star$, where $R_E$ is the Einstein radius  \citep{Wiktorowicz:21:SelfLensing}.
With a well-sampled pulse duration and amplitude and an estimate of the companion star mass and radius, we can eliminate the unobserved orbital period $P$ and derive the compact object mass.

To obtain the mass and radius of the companion star, we first cross-matched our pulse candidates to the stellar parameters catalog of \citet{2019ApJ...887...93G}, which provides an estimated absolute magnitude of the companion assuming it is on the main sequence.
We used the empirical luminosity--mass and mass-radius relations of \citet{Demircan:91:StellarMassRadiusLuminosity} to convert the absolute luminosity $L_\star$ into $M_\star$ and $R_\star$:
\begin{equation}
    M_\star = \left(\frac{L_\star}{1.18 L_\odot}\right)^{0.27} M_\odot
\end{equation}
and
\begin{equation}
    R_\star = 
    \begin{cases} 
1.06 R_\odot \left(\frac{M_\star}{M_\odot}\right)^{0.945}, & M_\star < 1.66 M_\odot \\
1.33 R_\odot \left(\frac{M_\star}{M_\odot}\right)^{0.555}, & M_\star > 1.66 M_\odot
    \end{cases}
\end{equation}

\begin{figure*}
    \centering
    \includegraphics[width=0.45\textwidth]{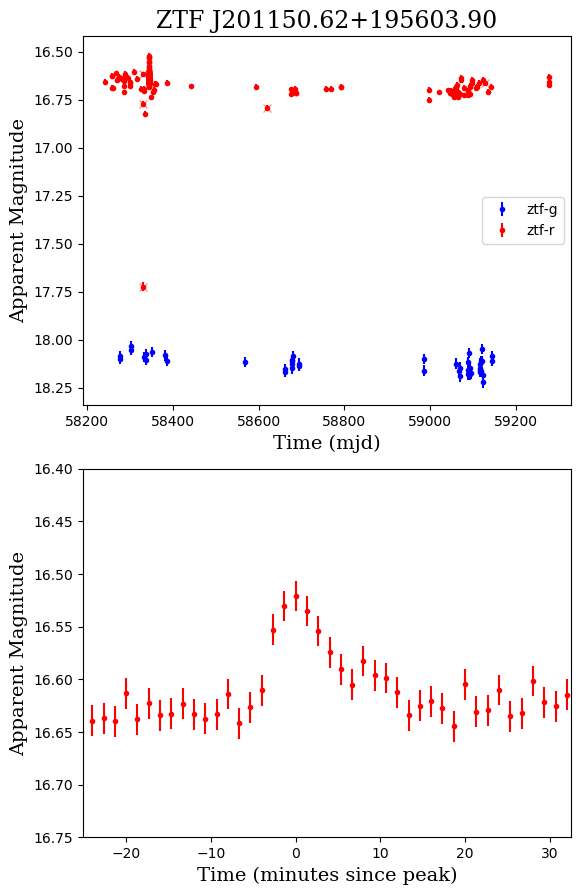}
    \includegraphics[width=0.44\textwidth]{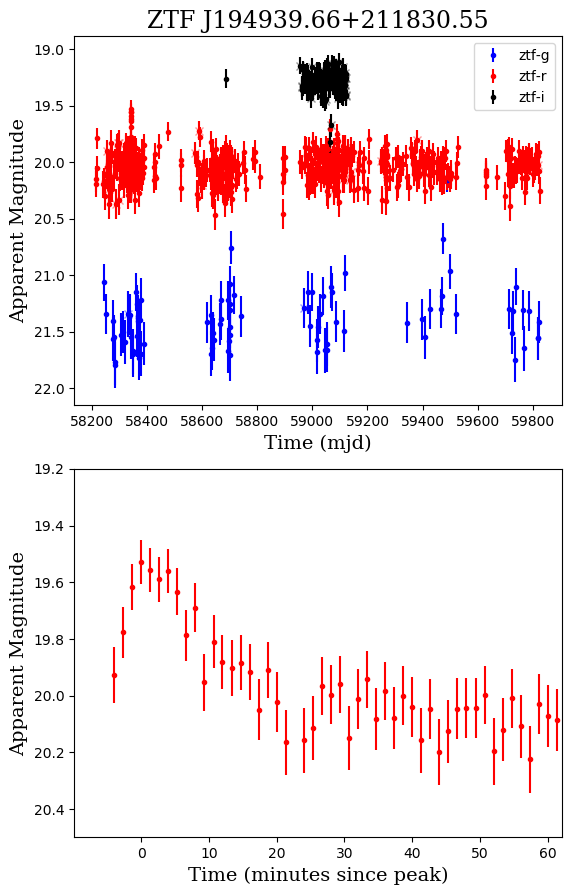}
    \caption{ZTF lightcurves of two self-lensing candidates with orbital periods not ruled out by ZTF observations (\S \ref{sec:Phase_Folding}). The top panel shows the full ZTF light curve history of the source while the bottom panel shows the identified pulse.} 
    \label{fig:final}
\end{figure*}

Given the coarseness of our assumptions about the companion star, these mass estimates are primarily useful in identifying pulses which would be implausible for stellar-mass compact objects.
We rule out four more objects based on improbable mass estimates, and are left with 19 candidates (Fig \ref{fig:final}). Table \ref{table:masses} provides the resulting mass estimates.

We evaluated whether the companion star would fill its Roche lobe given the binary component masses and orbital periods.
If so, we would expect temporal signatures of active accretion such as flickering that would mask the self-lensing signal and thus disfavor a self-lensing interpretation of the observed isolated pulse.
Only one candidate had a Roche filling factor larger than than one:
ZTF J185515.77$+$275426.63, with $R_\star/R_\mathrm{RL} = 1.46$.
This candidate is also ruled out by our periodicity search (\S \ref{sec:Phase_Folding}).

\subsection{Box Least Squares and Phase folding} \label{sec:Phase_Folding}

As a final step in the data analysis process, we looked for periodicity in the full ZTF light curve for each candidate. In addition to the small amount of continuous-cadence data from ZTF (\S \ref{sec:Data}), there is more sparsely sampled data spanning a much longer time range. Finding a periodic flux increase in the full, sparsely-sampled light curve would be strong evidence for a self-lensing origin. 
To search for this periodicity, we estimated the pulse period by computing a Box Least Squares Periodogram with the Astropy python package \citep{2022ApJ...935..167A} and phase folded the full light curve using the estimated period. 
No clear pulse period was determined for any of our sources using this method.

We attempted to determine whether the non-detection of periodicity could be used to rule out self-lensing as the explanation for these pulses, or whether ZTF did not have observations at the relevant periods.
Using the relations in \S \ref{sec:mass_estimation}, we derived best-fit orbital periods and uncertainties given the observed continuous-cadence pulse amplitudes, pulse durations, and inferred stellar masses and radii (Table \ref{table:masses}).
We constructed a grid of potential orbital periods ranging from plus to minus three times the standard deviation of the best-fit period\footnote{For one source, ZTF J194939.66+211830.55, our Monte Carlo estimate of the period uncertainty yielded a range too large for a practical search due to a larger uncertainty in its absolute magnitude.}.
We used period bins of one-third the pulse duration.
For each test period, we phase-folded the ZTF lightcurve omitting the night of continuous-cadence observations and counted the number of observations which fell into the phase bin defined by the start time and duration of the detected pulse.
If one or more ZTF observations occurred during the pulse phase, we ruled out self-lensing at that orbital period.
Dividing the number of periods not ruled out by the total number of test periods provides an estimate of the probability that the ZTF data support a self-consistent self-lensing explanation for the data (Table \ref{table:masses}).

Twelve systems remain plausible self-lensing candidates after this procedure (Appendix \ref{sec:app_a}).
However, most are unlikely to be due to self-lensing: only two candidates have more than 50\% of their potential orbital periods allowed by the ZTF non-detection of periodicity.
Seven of our nineteen self-lensing candidates (Appendix \ref{sec:app_b}) are ruled out entirely--ZTF would have detected self-lensing signatures at any of the potential orbital periods.

\section{Discussion} \label{sec:discussion}

From our initial dataset of 1910 light curves, we identify 12 candidate self-lensing pulses not caused by data artifacts, asteroid transits, or repeated stellar flares.
They also show reasonable estimates for the masses of the compact object and companion star (Table \ref{table:masses}). 
However, after analysis of all available ZTF data, they do not show a periodic brightening pulse as we would expect to see for a self-lensing system, and in most cases ZTF observations rule out periodic self-lensing pulses for the majority of plausible orbital periods.

We consider two other possible explanations for the observed brightening episodes.

The most probable explanation for these pulses is a stellar flare event. 
Stellar flares are extremely common, particularly on the low-mass stars that would produce the largest self-lensing signals \citep{tmp_Kowalski:24:StellarFlaresReview}.
Most stellar flares exhibit a characteristic light curve shape, with a fast rise and slow decay. 
While we have attempted to remove flares with the model selection procedure described in \S \ref{sec:model} and additional vetting of pulse shape  and outburst history (\S \ref{sec:Initial Screening}), many of our remaining candidates still exhibit a degree of asymmetry and out-of-pulse variability that suggest a flare origin.

Stellar flares exhibit a wide range of morphologies. 
Our stellar flare template, while derived from high-precision \textit{Kepler} data, may not be flexible enough to capture the full range of flare profiles.  
Additionally, studies of high-precision \textit{TESS} data suggest that as many as 9\% of stellar flares can be well-fit by Gaussian profiles \citep{Howard:22:NoSimpleFlares}.
Both effects may decrease the effectiveness of our model comparison procedure (\S \ref{sec:model}).

Lightcurve asymmetry and other flare characteristics can be identified with metrics in our continuous-cadence data but would be more challenging to discern at sparser cadences.
One possibility would be to assess the likelihood of observing a single large flare in our ZTF lightcurves using empirical flare frequency distributions for the inferred spectral type of each star.
However, the difficulty in conclusively determining whether a single pulse has a stellar flare origin suggests that periodicity searches will be more effective in confidently identifying self-lensing systems.

Isolated, short-duration pulses may also be due to microlensing by free-floating planetary-mass objects \citep[e.g.,][]{Sumi:11:FFPMicrolensing, Sumi:23:FFPMicrolensing, Mroz:17:FFPMicrolensing,Mroz:18:NeptuneFFP}.
In such cases the unknown distance between the background source and the lens is not negligible, and the lens mass cannot be determined directly due to degeneracies between the lens mass, lens distance, and source-lens relative proper motion.
Microlensing events due to lensing by terrestrial-mass objects with short durations comparable to these ZTF pulses have been reported \citep[e.g.,][]{Mroz:20:EarthFFP}.
In such cases, finite-source effects may be detected \citep[e.g.,][]{Witt:94:FiniteSourceEffects}, improving constraints on the mass of the lens by eliminating the degeneracy due to source-lens relative proper motion.
However, inference of the underlying lens mass distribution still relies on statistical samples.
Distinguishing these events from stellar flares is likewise challenging.
We defer detailed analysis and microlensing model fitting of these events to future work.

We roughly estimate the number of self-lensing systems expected in our pilot search as follows.
\citet{Wiktorowicz:21:SelfLensing} estimate $1.8\times10^3$ self-lensing systems could be detected in at least one epoch in ZTF assuming a standard IMF.  
Their calculation assumes a five-year ZTF survey with a simplified 1-day cadence.
Crossmatching ZTF Data Release 14 to the PanSTARRS catalog \citep{Flewelling:16:PS1db}, we find 1,234,463,018 unique ZTF lightcurves.
This pilot search considered 117,483,764 lightcurves over 14 days.
Scaling by the fraction of lightcurve-days searched, we would expect 1.3 self-lensing events in this continuous-cadence dataset.
Thus while this pilot search was sufficiently large that observing self-lensing was possible, our nondetection of confirmed self-lensing events is also plausible.

\section{Conclusion} \label{sec:Conclusion}

In this pilot search for non-interacting Galactic compact binaries with the Zwicky Transient Facility, we searched for self-lensing signatures in order to detect these otherwise hidden objects. 
Although we found no clear-cut cases of periodic self-lensing signatures in this initial dataset, this study lays the groundwork for larger searches of the complete ZTF lightcurves.
Population synthesis suggests thousands of detached compact binaries could be discovered in ZTF \citep{Wiktorowicz:21:SelfLensing}, although our study highlights that a range of contaminants, particularly image artifacts, stellar flares, and microlensing by free-floating planets, will complicate such a search.
Nevertheless, a systematic search of the ZTF data would provide an opportunity to probe an otherwise inaccessible compact binary population and improve our understanding of binary and stellar evolution.

\section*{Acknowledgements}

We thank Eric Agol and Matthew Middleton for useful discussions.

Support for this work was provided in part by the Research Corporation Scialog grant \#24221. 
ECB acknowledges support by the NSF AAG grant 1812779 and grant \#2018-0908 from the Heising-Simons Foundation.

This research has made use of IMCCE's SkyBoT VO tool.

This research has made use of the NASA/IPAC Infrared Science Archive, which is funded by the National Aeronautics and Space Administration and operated by the California Institute of Technology.

Based on observations obtained with the Samuel Oschin Telescope 48-inch and the 60-inch Telescope at the Palomar Observatory as part of the Zwicky Transient Facility project. ZTF is supported by the National Science Foundation under Grant No. AST-2034437 and a collaboration including Caltech, IPAC, the Weizmann Institute of Science, the Oskar Klein Center at Stockholm University, the University of Maryland, Deutsches Elektronen-Synchrotron and Humboldt University, the TANGO Consortium of Taiwan, the University of Wisconsin at Milwaukee, Trinity College Dublin, Lawrence Livermore National Laboratories, and IN2P3, France. Operations are conducted by COO, IPAC, and UW. 

\bibliographystyle{yahapj}
\bibliography{self_lensing}


\appendix
\section{Appendix A} \label{sec:app_a}

ZTF lightcurves of 12 plausible self-lensing candidates. 
For each source, the top panel shows the full ZTF lightcurve and the lower is zoomed on the pulse detected in the $r$-band continuous-cadence data.  Data points with nonzero pipeline flags are marked with a x and excluded from analysis.

\includegraphics[width=0.4\textwidth]{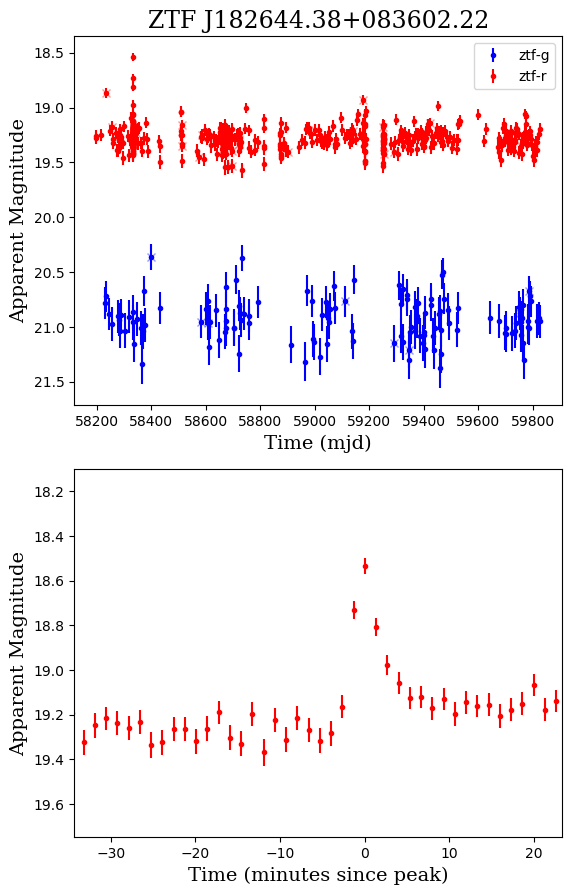}
\includegraphics[width=0.4\textwidth]{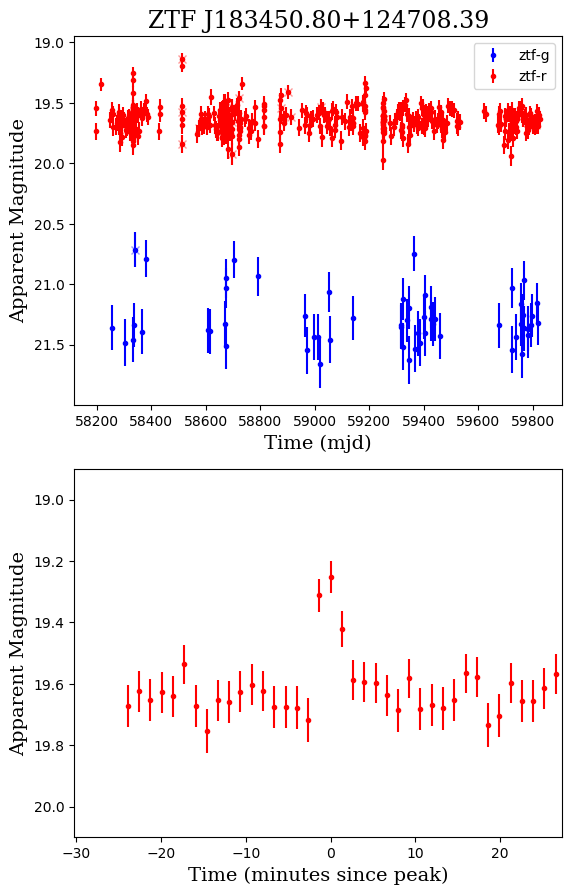} 
\includegraphics[width=0.4\textwidth]{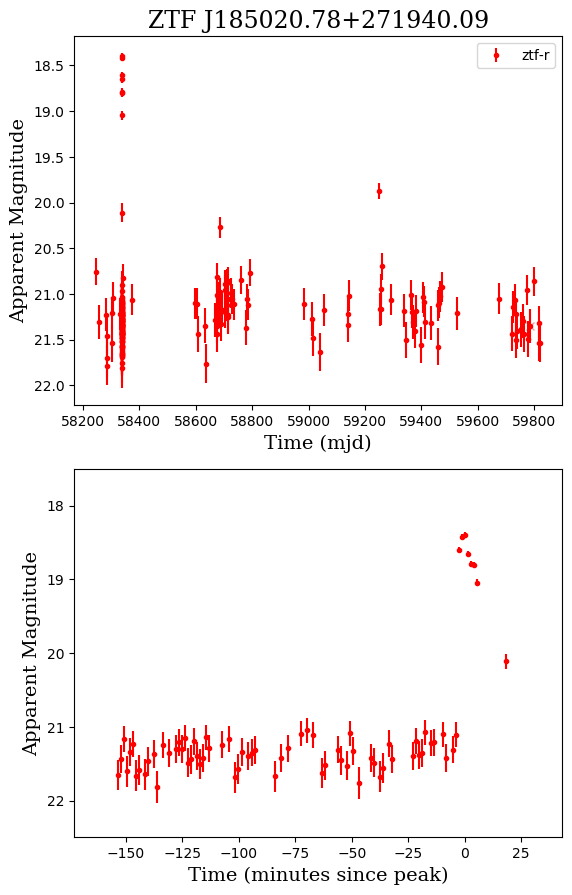} 
\includegraphics[width=0.4\textwidth]{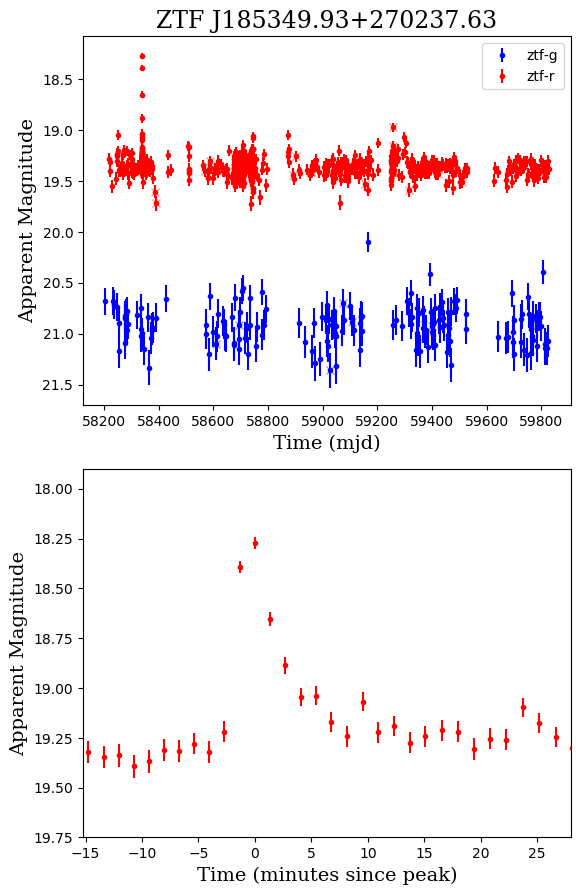} 
\includegraphics[width=0.4\textwidth]{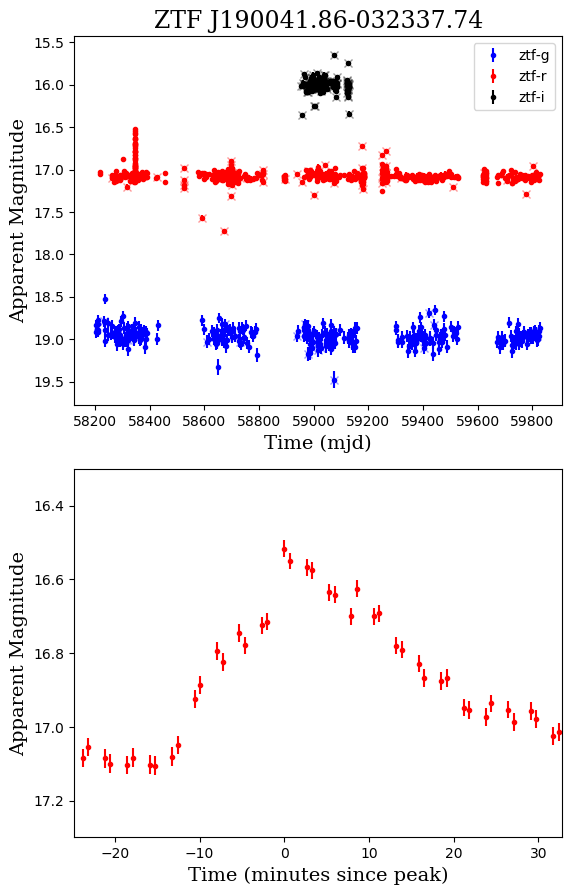} 
\includegraphics[width=0.4\textwidth]{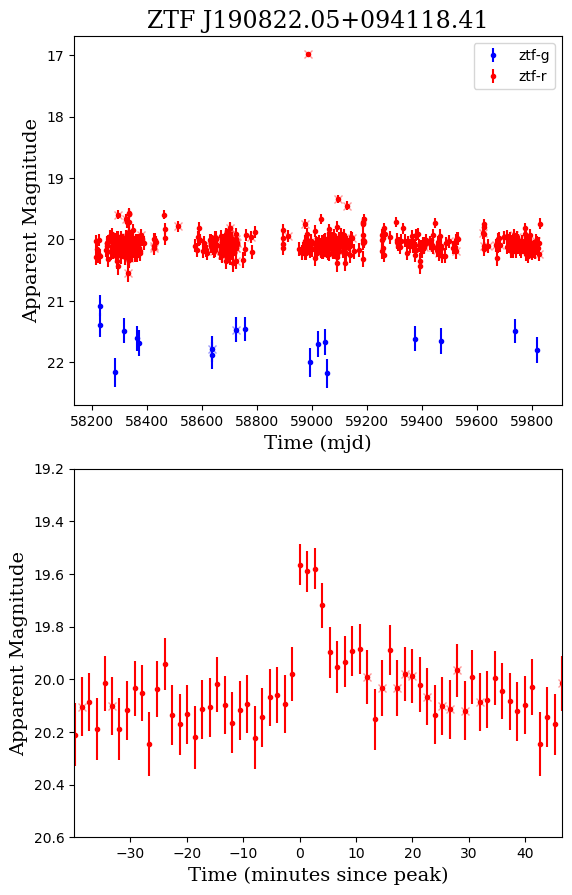} 
\includegraphics[width=0.4\textwidth]{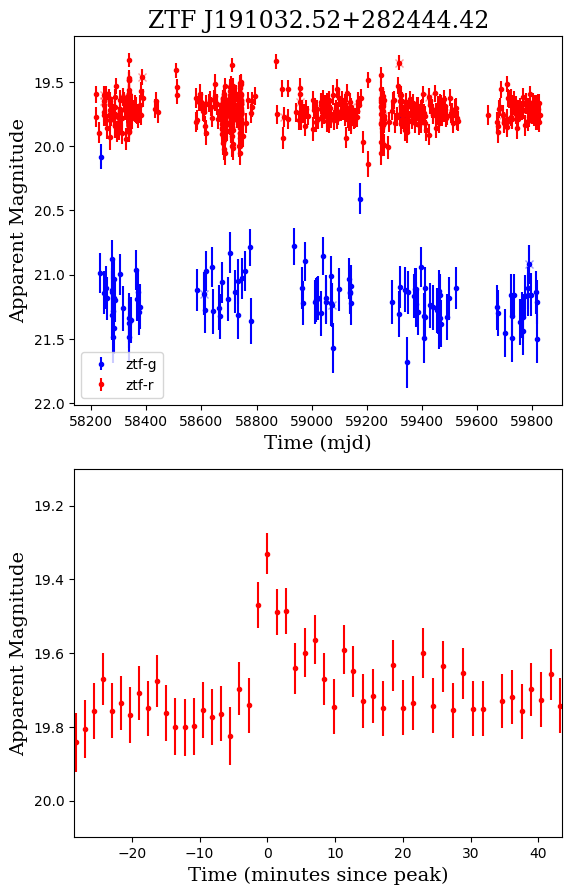} 
\includegraphics[width=0.4\textwidth]{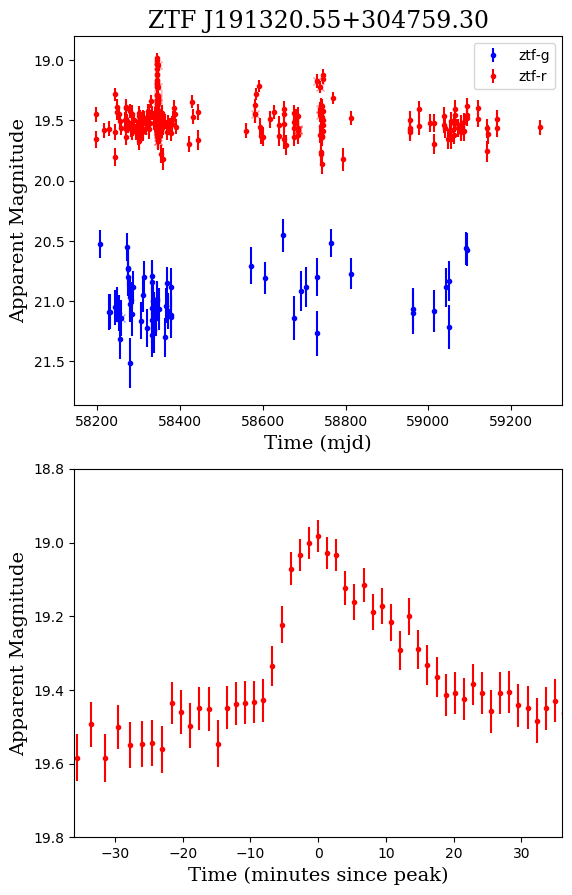} 
\includegraphics[width=0.4\textwidth]{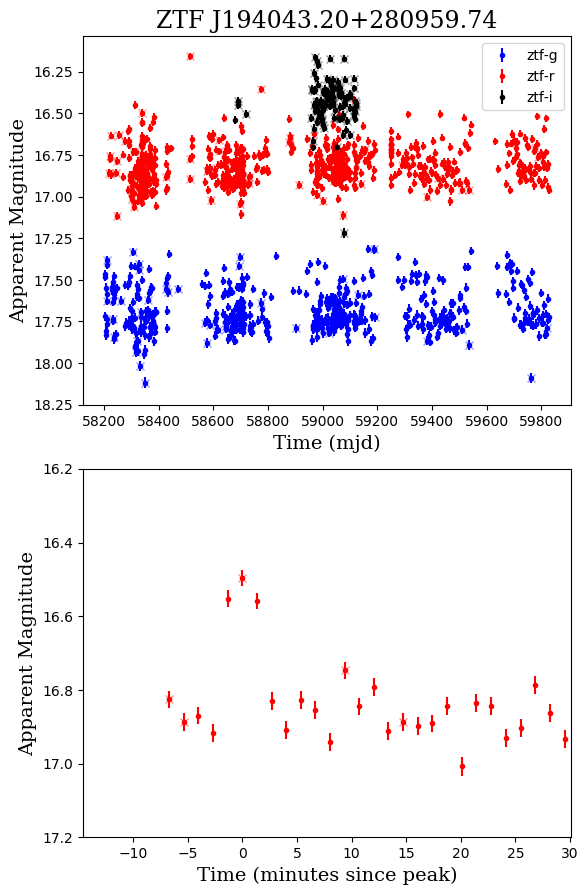} 
\includegraphics[width=0.4\textwidth]{Images/194_fig.png} 
\includegraphics[width=0.4\textwidth]{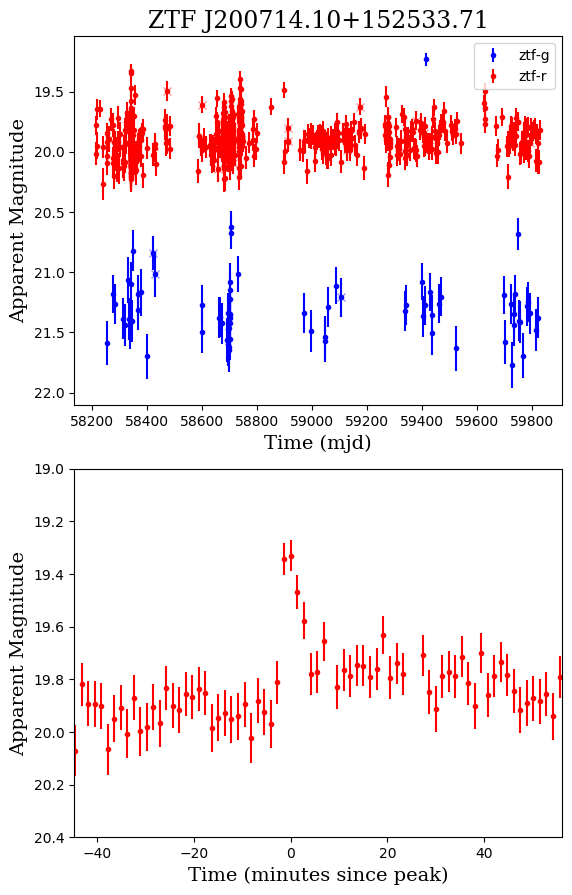} 
\includegraphics[width=0.4\textwidth]{Images/2011_fig.png} 

\section{Appendix B} \label{sec:app_b}
ZTF lightcurves of 7 sources for which a self-lensing explanation is ruled out by period analysis (\S \ref{sec:Phase_Folding}). 
Plots as in Appendix \ref{sec:app_a}.

\includegraphics[width=0.4\textwidth]{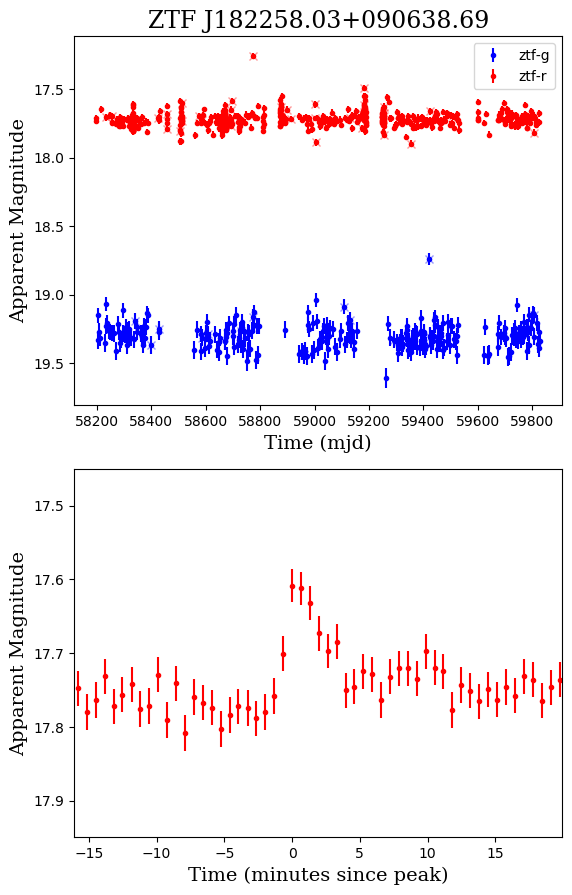}
\includegraphics[width=0.4\textwidth]{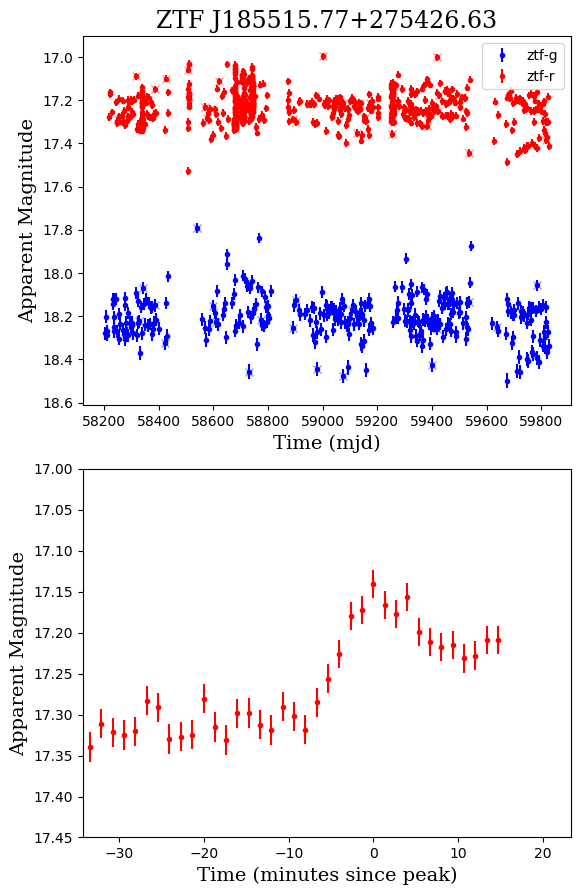} 
\includegraphics[width=0.4\textwidth]{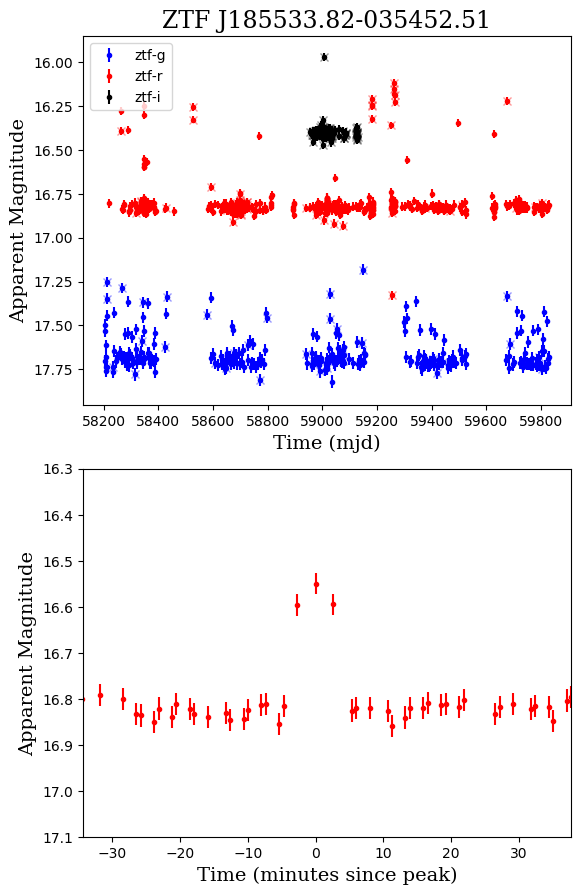}
\includegraphics[width=0.4\textwidth]{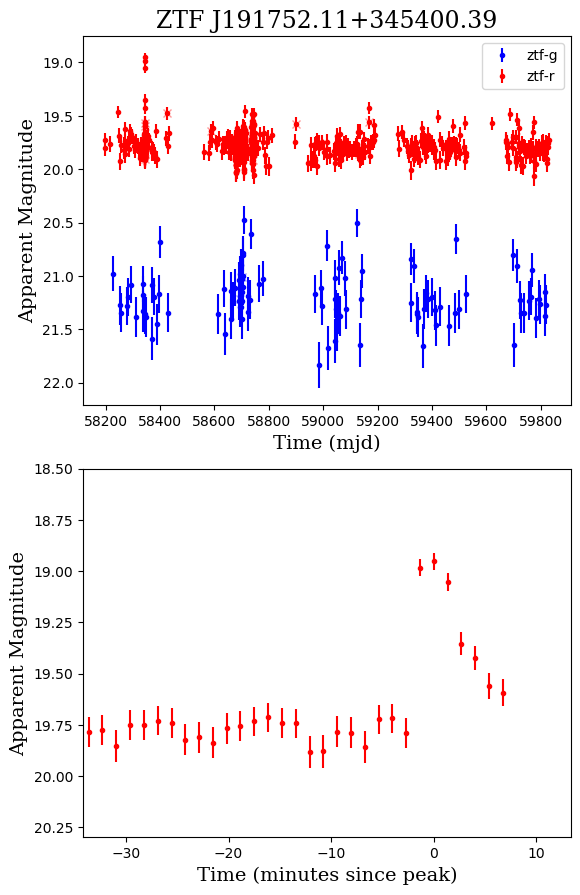}
\includegraphics[width=0.4\textwidth]{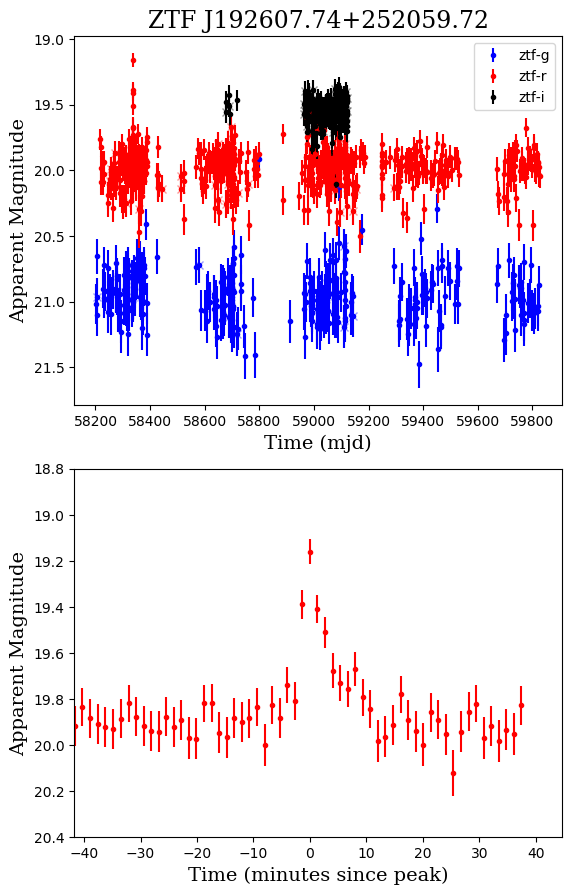}
\includegraphics[width=0.4\textwidth]{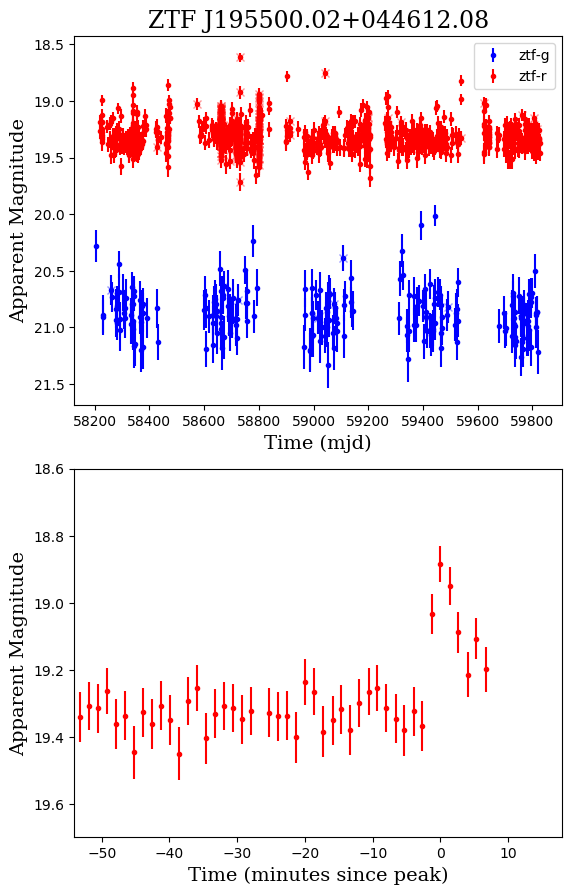}
\includegraphics[width=0.4\textwidth]{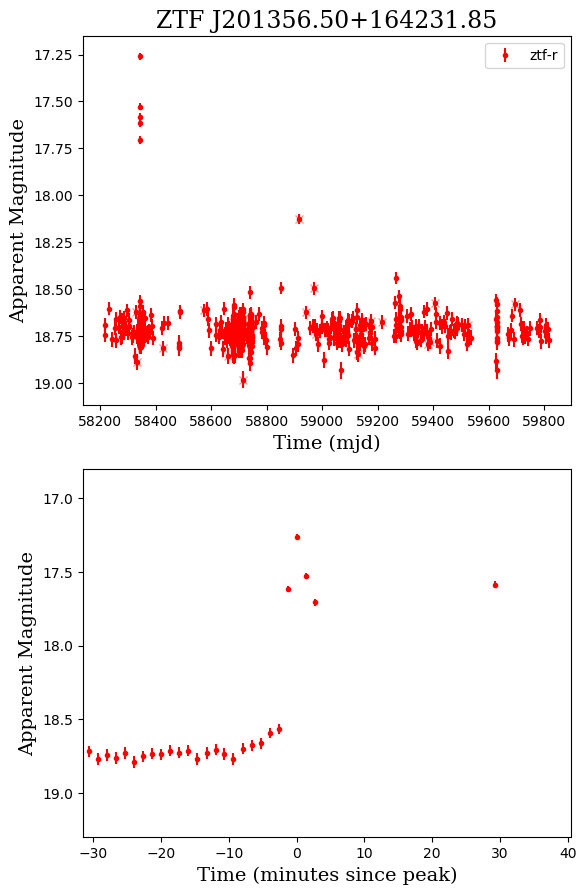} 


\begin{sidewaystable}
    \centering
    \begin{tabular}{|c|c|c|c|c|c|c|c|c|c|c|c|c|c|c|}
    \hline
    Name & Absolute & Pulse & Peak Time & Star Mass & Star Radius & Compact Object & Inferred Period & Fraction \\
    & Magnitude & Duration & (MJD) & ($M_\odot$) & ($R_\odot$) & Mass ($M_\odot$) & (days) & Periods \\
    &&(minutes)&&&&&& Allowed\\
    \hline

    ZTF J182644.38+083602.22 & 11.35\spm{0.30}{0.25} & 23.9 & 58334.27 & $0.18\pm0.01$ & $0.22\pm0.01$ & $1.63\pm0.24$ & $4.09\pm0.32$ & 0.13 \\ \hline
    ZTF J183450.80+124708.39 & 13.10\spm{0.20}{0.25} & 25.3 & 58334.21 & $0.12\pm0.01$ & $0.14\pm0.01$ & $0.51\pm0.07$ & $5.72\pm0.43$ & 0.31 \\ \hline
    ZTF J185020.78+271940.09 & ... & 13.4 & 58337.32 & ... & ... & ... & ... & ...\\ \hline
    ZTF J185349.93+270237.63 & 10.55\spm{0.25}{0.20} & 26.3 & 58338.23 & $0.23\pm0.01$ & $0.26\pm0.02$ & $2.73\pm0.33$ & $5.03\pm0.32$ & 0.11 \\ \hline
    ZTF J190041.86-032337.74 & 6.62\spm{0.38}{0.38} & 52.9 & 58348.31 & $0.60\pm0.06$ & $0.65\pm0.06$ & $10.36\pm1.95$ & $9.53\pm0.97$ & 0.05 \\ \hline
    ZTF J190822.05+094118.41 & 10.25\spm{0.60}{0.47} & 41.2 & 58334.23 & $0.24\pm0.03$ & $0.28\pm0.04$ & $1.99\pm0.56$ & $11.81\pm1.98$ & 0.32 \\ \hline
    ZTF J191032.52+282444.42 & 10.08\spm{0.37}{0.43} & 34.3 & 58337.24 & $0.25\pm0.03$ & $0.29\pm0.03$ & $2.02\pm0.47$ & $6.20\pm0.79$ & 0.14 \\ \hline
    ZTF J191320.55+304759.30 & 10.05\spm{0.30}{0.25} & 64.6 & 58345.27 & $0.26\pm0.02$ & $0.29\pm0.02$ & $1.83\pm0.27$ & $37.01\pm2.97$ & 0.83 \\ \hline
    ZTF J194043.20+280959.74 & ... & 21.4 & 58337.22 & ... & ... & ... & ... & ... \\ \hline
    ZTF J194939.66+211830.55 & 7.00\spm{1.50}{2.17} & 48.0 & 58343.20 & $0.55\pm0.39$ & $0.60\pm0.40$ & $8.50\pm12.87$ & $7.65^\dagger$ & ... \\ \hline
    ZTF J200714.10+152533.71 & 11.28\spm{0.42}{0.38} & 32.7 & 58344.26 & $0.19\pm0.02$ & $0.22\pm0.02$ & $1.32\pm0.27$ & $8.30\pm0.95$ & 0.27 \\ \hline
    ZTF J201150.62+195603.90 & 9.25\spm{0.10}{0.10} & 56.3 & 58343.24 & $0.31\pm0.01$ & $0.35\pm0.01$ & $1.54\pm0.08$ & $12.40\pm0.36$ & 0.74 \\ \hline \hline
    ZTF J182258.03+090638.69 & 8.80\spm{0.15}{0.10} & 18.4 & 58333.26 & $0.35\pm0.01$ & $0.39\pm0.01$ & $2.93\pm0.21$ & $0.56\pm0.02$ & 0.0 \\ \hline
    ZTF J185515.77+275426.63 & 2.35\spm{0.47}{0.75} & 28.1 & 58338.31 & $1.74\pm0.36$ & $1.81\pm0.21$ & $59.50\pm14.08$ & $0.38\pm0.05$ & 0.0 \\ \hline
    ZTF J185533.82-035452.51 & 2.55\spm{0.50}{0.60} & 50.3 & 58348.25 & $1.65\pm0.27$ & $1.70\pm0.20$ & $56.09\pm13.72$ & $2.44\pm0.32$ & 0.0 \\ \hline
    ZTF J191752.11+345400.39 & 11.25\spm{0.37}{0.40} & 12.1 & 58345.30 & $0.19\pm0.02$ & $0.22\pm0.02$ & $2.26\pm0.46$ & $0.67\pm0.08$ & 0.0 \\ \hline
    ZTF J192607.74+252059.72 & 5.80\spm{1.10}{0.90} & 29.4 & 58338.30 & $0.74\pm0.18$ & $0.79\pm0.19$ & $20.80\pm10.17$ & $1.80\pm0.82$ & 0.0 \\ \hline
    ZTF J195500.02+044612.08 & 10.70\spm{0.25}{0.25} & 10.6 & 58340.30 & $0.22\pm0.01$ & $0.25\pm0.02$ & $2.28\pm0.29$ & $0.31\pm0.02$ & 0.0 \\ \hline
    ZTF J201356.50+164231.85 & 10.25\spm{0.20}{0.20} & 6.7 & 58343.30 & $0.24\pm0.01$ & $0.28\pm0.01$ & $6.12\pm0.60$ & $0.14\pm0.01$ & 0.0 \\ \hline
    
    \end{tabular}
    \caption{ZTF self-lensing candidates identified by their sexigesimal RA and Dec. Absolute magnitude estimates are taken from \citet{2019ApJ...887...93G}; two candidates were not cataloged. $^\dagger$The orbital period of ZTF J194939.66+211830.55 was poorly constrained.} 
    \label{table:masses}
\end{sidewaystable}

\end{document}